\documentstyle[12pt]{article}

\newcommand{\be}{\begin{equation}}
\newcommand{\ee}{\end{equation}}
\newcommand{\bea}{\begin{eqnarray}}
\newcommand{\eea}{\end{eqnarray}}

\begin{document}
\begin{titlepage}


\vspace{1in}

\begin{center}
\Large
{\bf NSR STRING IN ${\bf {AdS_3 }}$ BACKGROUND: NONLOCAL CHARGES}

\vspace{1in}

\normalsize

\large{  Jnanadeva  Maharana\\
E-mail maharana$@$iopb.res.in 
  }

\normalsize
\vspace{.5in}

 {\em Institute of Physics \\
Bhubaneswar - 751005 \\
India  \\ }

\end{center}

\vspace{1in}

\baselineskip=24pt
\begin{abstract}
We consider NSR superstring in $AdS_3\otimes S^3\otimes R_4$ background. 
The action
is expressed in terms of the variables on the group manifolds $SL(2,R)$ and
$SU(2)$ as $1+1$ dimensional $\sigma$-models with Wess-Zumino terms,
associated with 
$AdS_3$ and $S^3$ respectively. $R_4$ is flat Euclidean space. We show
the existence of classical nonlocal conserved currents in the superspace
formulation for the nonlinear $\sigma$-model with WZ term in the context
of NSR string. We propose Ward identities utilizing  existence of 
the nonlocal conserved
currents.

\end{abstract}

\vspace{.7in}
 
\end{titlepage}



The Maldacena conjecture establishing AdS/CTF correspondence has stimulated
activities in diverse directions \cite{mald,kleb,witt,rev}.  There have been considerable amount of interests in the study of integrability
properties of evolution of strings in AdS backgrounds.
The pp-wave limit of $AdS_5$ geometry \cite{bmn} facilitates the derivation
of the string spectrum in that background and therefore, a string theoretic
test of the AdS/CFT correspondence could be realized. Subsequently,
 the quantization of string theory in the semiclassical approach was 
undertaken which might be interpreted as going beyond the BMN limit. Here,
the quantization of the $\sigma$-model is carried out in the vicinity of a 
classical solution \cite{gkp,ark}; whereas BMN used BPS null geodesics. 
  Thus intuitively, it is
tempting to conjecture that strings in AdS background might be intimately
related to integrable systems.\\
We recall that the $AdS_5 \otimes S^5$ geometry emerges from the type IIB
theory as follows: one considers $D1 - D5$ brane configurations in 
10-dimensions which is suitably compactified to obtain a five dimensional BPS
black hole solution. Then, the near horizon limit is taken and the resulting 
geometry is identified to be $AdS_5 \otimes S^5$. The seminal results of 
Maldacena  are based on realization of the afore mentioned
 geometry for the $D1 -D5$ system. Note that the superstring propagates
in the $AdS_5 \otimes S^5$ background metric in the presence of constant
5-form RR fluxes. The NSR formulation is inadequate to construct a suitable 
nonlinear  sigma model for type IIB superstring in such backgrounds. 
It is
argued that the Green-Schwarz formulation is more convenient to construct
the desired $\sigma$-model action for the type IIB theory.  However, the
action is constructed in the light-cone gauge and there are several proposals
for construction of the action \cite{mt,kt,rs}. Bena Roiban and Polchinski 
constructed nonlocal currents for type IIB
theory when the $\sigma$-model is described on the
coset ${{PSU(2,2!4)}\over {SO(1,4)\otimes SO(5)}}$. 
Subsequently, a lot of
attention has been focused in this direction. Recently, explicit construction
of the nonlocal charges have been presented \cite{hy} by adopting the
Roiban-Seigel form \cite{rs} of the action for a careful handling of the 
Wess-Zumino (WZ) term that arises for type IIB $\sigma$-model on $AdS_5 \otimes
S_5$ geometry. In another development, the integrability of type IIB superstring
in AdS background has been investigated in the Hamiltonian formalism 
for the sigma model action in
\cite{dmms}. The transition matrices are defined and their Poisson bracket 
algebra derived. The existence of infinite set of nonlocal currents,
 in the bosonic
sector,  for strings in $AdS_5
\otimes S^5$, which is described by a nonlinear $\sigma$-model,
 was constructed in \cite{msw}. Therefore,  it provided an evidence that the
theory is classically integrable. 
 \\
Polyakov \cite{pkov} has explored gauge/string
correspondence from another perspective for noncritical string theories. 
The proposal is to envisage
$\sigma$-models on $AdS_p \otimes S^q$, $p+q <10$. He has argued that
such models are conformal and are integrable, although quantum integrability
associated with the conserved charges is not yet proved \cite{pkov}. These
models correspond to conformal fixed points of gauge theories in diverse
dimensions.\\
The existence of Yangian symmetries have been shown by Dolan, Nappi and Witten
\cite{dnw} for the ${\cal N }=4 $ SYM in the limit of vanishing coupling
constant. They argue that the Yangian symmetry observed on the SYM side
might be intimately connected to similar symmetry on the string theory
side due the presence of conserved nonlocal charges obtained in
\cite{bpr} from the string
worldsheet perspective. Therefore, the study of integrability properties of
strings in AdS background is expected to unravel another aspect of the 
AdS/CFT correspondence.\\
The purpose of this article is to construct family of nonlocal conserved
currents for type IIB superstring on $AdS_3\otimes \otimes  S^3
\otimes {\cal C}_4$, 
where ${\cal C}_4$ is the four dimensional flat  space; a preliminary
version of the work has been  presented a while ago \cite{bjm}. Thus
we are studying a superstring in the NS-NS sector in 10-dimensions.
The type IIB theory
admits both NS-NS and RR 3-form antisymmetric field strengths for the $S^3
\otimes AdS_3$ geometry. Of course, the field strength is required to be
covariantly constant. We confine our attention to the NS-NS background.
The antisymmetric two form NS-NS potential
is set to zero on the ${\cal C}_4$.
Therefore, the goal is to construct an appropriate $\sigma$-model action for the
string in the $S^3\otimes AdS_3$ configuration. It is worth while to record 
a few comments at this stage before writing the action explicitly.
  The general form of the  worldsheet action for the NSR string as a  
$\sigma$-model action in backgrounds of target space metric and torsion
is is constructed
as follows
\bea
S=\int d^2 \sigma d {\theta}  d{\bar  \theta}\left( {1\over 2} 
G_{MN}({\hat Z}) 
{\bar D}{\hat Z}^M D{\hat Z}^{N}+{1\over 2}B_{MN}({\hat Z}) {\bar D} 
{\hat Z}^MD{\hat Z}^{N} \right)
\eea
where $M,N=0,1,2...9$ are the spacetime indices of string coordinates. The
worldsheet coordinates are denoted as $\tau$ and $\sigma$ and 
${\bar \theta}, \theta  $
are  Grassmann variables defining the superspace. The super covariant
derivatives are defined as
\bea
D_a={{\partial}\over {\partial {\bar \theta} ^{a}}}-i(\gamma ^{\alpha}
\theta)_a 
\partial _{\alpha}
\eea
The superfield ${\hat Z}^M$ has the bosonic field as its first component and is
expressed as 
\be
{\hat Z}^{M}=Z^M+i{\bar \theta} \psi ^M +{i\over 2}{\bar \theta}
\theta  F^M
\ee
The chiral fermions $\psi ^M$  are the super partners of bosonic coordinates
and $F^M$ is the auxiliary field. We expand the backgrounds which are functions
of the super fields and note that $\int d\theta  \theta  =1$ and $\int d{\bar
\theta}{\bar \theta} =1$. 
Thus only finite number of terms contribute to the Grassmann integrals in
the expansion. We recall \cite{abdus,dmr}
 that besides the usual quadratic bosonic 
and fermionic kinetic energy terms, there are also generic quadratic and 
quartic  fermionic couplings  involving 
the 
Riemann curvature tensor: $R_{MNRS}{\bar \psi}^M \psi
^R{\bar \psi}^N\psi ^S$,   a  term involving the field strength of $B_{MN}$:
$H_{MNP}{\bar \psi}^M\gamma^{\alpha}\gamma _{5}\psi ^N\partial _ 
{\alpha}{\hat Z}^P$ and another four fermion interaction term 
${\cal D}_M H_{NPR}
{\bar \psi}^N\psi ^M{\bar {\psi}}^R\psi ^P$, where ${\cal D}_M$ is the 
covariant derivative defined with the Christofel connection of the target
space metric.\\
Let us  consider propagation of the NSR string on 
the geometry $AdS_3 \otimes S^3 \otimes R_4 $, where $R_4$ is flat Euclidean
space and the antisymmetric B-field has vanishing components along $R_4$.  This 
  background configuration is adopted so that
 the  resulting action takes 
a simple form.
  We denote the coordinates on $AdS_3$ as $X^{\mu}, \mu=0,1,2$;
those on $S^3$ as $Y^I, ~I=1,2,3$. The four coordinates of $R_4$ are $Z^6, Z^7,
Z^8 ~{\rm and}~ Z^9$; $R_4$. It is well known that the
action for propagation of NSR superstring  on $AdS_3$ background  can be
expressed as a supersymmetric Wess-Zumino-Witten model on $SL(2,R)$ group
manifold. Similarly, the $S^3$ part of the action is a WZW action on $SU(2)$
group manifold. The bosonic part of $AdS_3$ coordinates are expressed an
an element of $SL(2,R) $ 
\bea
\label{sl2r}
{\tilde g}= \pmatrix { X_{-1}+X_1  & X_0-X_2  \cr -X_0-X_2  & X_{-1}-X_1 \cr}
\eea
with  the constraint $ X_{-1}^2+X_0^2
-X_1^2-X_2^2=1 $. We introduce an extra coordinate, $X_{-1}$, to  define the
group element $\tilde g$, satisfying ${\rm det}{\tilde g}=1$.
The analogous $SU(2)$ matrix is
\bea
\label{su2}
g=\pmatrix { Y_1+iY_4 & Y_2+iY_3 \cr -(Y_2-iY_3) &Y_1-iY_4 \cr}
\eea
satisfying the constraint is $Y_4^2+Y_1^2+Y_2^2+Y_3^2=1 $ and $Y_4$ is added
to define $SU(2)$ element $g$. The worldsheet supersymmetric can be expressed 
 in terms of the set of coordinates
$\bf X$ and $\bf Y$ which define the matrices $\tilde g$ and $g$ and introduce
the worldsheet fermions as alluded to earlier. Then, we could follow
the prescriptions for construction of nonlocal currents as given by Curtright
and Zachos \cite{cs}. The same techniques have been generalized to 
construct infinite set of nonlocal conserved currents for supersymmetric 
nonlinear $\sigma$-models defined Grassmann manifolds \cite{jm}. 
The Poisson bracket algebra of 
the corresponding nonlocal charges could be computed adopting the Direc
constrained Hamiltonian formalism \cite{adjm}. 
Moreover, as
has been demonstrated  \cite{cs,jm}, when we deal separately with
currents constructed as bilinear in bosonic fields and bilinear in the
fermionic fields, these are not separately conserved, but their divergences
appear as products of the currents. Nevertheless, the sum of the two
currents is conserved on shell. As is well known, the curl of each of the
currents and their linear combinations do not vanish. Consequently, the 
procedure due to Br\'ezin, Itzykson, Zinn-Justin and Zuber \cite{chiral} is 
found to be inadequate. A more elegant and economic procedure is to appeal
directly to superspace formulation to construct the nonlocal
currents \cite{chau}. Therefore, the goal is
to write the supersymmetric action on a group manifold in the superspace.
The procedure for  construction of such actions is well known
 \cite{abd,paolo,bcz}.\\
We now proceed to remind the reader how  the bosonic part of two 
dimensional WZW model is defined on a group manifold. Let us consider the  $SU(2)$
group (the form of the action below is also valid for  $SU(N)$ group). 
The case of $SL(2,R) $ could be treated
in parallel and we shall  point out how  the construction of the action
for  $SL(2,R)$ will differ from $SU(2)$.
\bea
\label{bwzw}
S_B={1\over {4\lambda ^2}}\int d^2\sigma {\rm  Tr}\left(\partial _{\alpha}
g^{-1}(\sigma ,\tau) \partial ^{\alpha }g(\sigma , \tau) \right) +
{k\over{16\pi }}
\int _B{\rm Tr}\left(g^{-1}dg\wedge g^{-1}dg \wedge g^{-1}dg\right)
\eea
Here $\lambda$ and $k$ are  dimensionless coupling constants; 
for compact group like $SU(2)$,
the coefficient appearing in front of WZ term, $k$ is quantized
 for the consistency of the quantum theory. $g(\sigma ,
\tau)$ takes values in $SU(2)$. Note that $g$ should
be extended smoothly to a 3-dimensional manifold, B and its boundary is our
world sheet (actually one should  define complex coordinates in
the standard manner and write the action in those variables) which enables us
to write the WZ term this way. The action is invariant under
global $SU(2)_L \otimes SU(2)_R$ transformations:  $g\rightarrow
{\cal W}_L g{\cal W}_R $ where ${\cal W}_{L, R} \in SU(2)_{L, R}$. We remark
in passing that the WZ term, expressed in this form (\ref{bwzw}) makes the
action manifestly invariant under the above global symmetry. \\
We note the following  few features of the noncompact group
$SL(2,R)$ the counter part $SU(2)$ which is relevant as well:
(i) a matrix  ${\tilde g}\in SL(2,R)$ satisfies 
${\tilde g}^T \eta {\tilde g}=\eta$,
contrast this constraint  with $g\in SU(2)$ satisfying $gg^{\dagger}=1$.
Here $\eta$ is the $SL(2,R)$ metric, with the property that $\eta ^2=-{\bf 1}$
and it can be chosen to be 
\bea
\label{slr}
\eta=\pmatrix {0 & 1 \cr -1 & 0 \cr}
\eea
Therefore, this property of $\tilde g$ should be taken into account in future
considerations. (ii) Due to its noncompact nature, for $SL(2,R)$ 
the coefficient appearing in front of WZ term is not quantized. So long as we are 
dealing only with the classical theory $k$ need not be quantized for
the  for $SU(2)$ also.\\
Let us focus on the $SU(2)$ from now on.  The WZW action in the \cite{abd,paolo}
 superspace is
\bea
\label{action}
S={1\over {4\lambda ^2}}\int d\sigma d\tau d\theta d{\bar \theta}
{\bar D}G^{\dagger}DG +{k\over {16\pi}}\int d\sigma d\tau d\theta d{\bar \theta}
\int _0^1dt G^{\dagger}{{dG}\over {dt}}{\bar D}G^{\dagger}\gamma _5DG
\eea
$G$ is the $SU(2)$ matrix superfield, satisfying  $G^{\dagger}G
=\bf 1$. In order to define the WZ term as an integral over a three 
dimensional space  one defines an extension of the superfield to 
 3 dimensions
so that $t=0$ corresponds to the superfield on the two dimensional space.
Here $\gamma _5 =\sigma _3$ is the two dimensional $\gamma _5$ matrix. In the 
WZW theory $\lambda ^2 ={4\pi \over k}$; corresponds to the point where
the theory becomes conformally invariant. 
When this condition is not fulfilled there
are quartic fermionic couplings in this action too after we eliminate the
auxiliary fields. 
 Notice that for propagation of the string we must enforce conformal
invariance and therefore the above relation between couplings must hold.
Consequently, the quartic couplings involving fermions are absent.\\
It is more convenient to express the currents and their conservation laws in 
terms light cone variables. Another advantage is that  in this frame, 
we write the currents
in terms of chiral fermions which considerably simplify all computations. 
 Furthermore, the equations of motion take 
simple form of current conservations.\\
Let us define the $SU(2)$ currents and discuss their conservation laws
which will be the basis for the construction the infinite set of
 nonlocal conserved
currents. The matrix superfield is defined (in light cone variables) as
\bea
\label{superf}
G(\sigma ,\tau ,\theta)=g(\sigma ,\tau)\left({\bf 1}+i\theta ^+\psi _+(\sigma ,
\tau )+i\theta ^-\psi _-(\sigma ,\tau)+i\theta ^+\theta ^-F(\sigma ,\tau) 
\right)
\eea
Correspondingly, the supercovariant derivatives are
\bea
D_{\pm}={{\partial}\over {\partial {\theta}^{\pm}}}-i\theta ^{\pm}
\partial _{\pm}~~~{\rm with}~~ {\partial}_{\pm}=\partial _{\tau}\pm\partial _
{\sigma}
\eea
the light cone variables are defined as $\sigma ^{\pm} ={1\over 2}(\tau \pm
\sigma)$

The chiral fermions $\psi _{\pm}$ are matrices take value in the Lie algebra 
and F is the auxiliary field. 
The condition $G^{\dagger}G={\bf 1}$ on the superfield translates 
into constraints on  the component fields  $g$,  $\psi _{\pm}$ and $F$.
The action can be obtained in terms of component
fields \cite{abd,paolo} after eliminating F.
Note the  following properties of  $g(\sigma ,\tau)$ and 
$\psi _{\pm}(\sigma ,\tau)$ under the global $SU(2)_L \otimes SU(2)_R$
transformations
\bea g\rightarrow {\cal W}_Lg{\cal W}_R ^{-1} ~~~~ {\rm and}~~~~
\psi _{\pm}\rightarrow
{\cal W}_R\psi _{\pm}{\cal W}_R^{-1},
\eea
where ${\cal W}_L\in SU(2)_L$ and ${\cal W}_R \in SU(2)_R$.
Let us introduce the supercurrent
\bea
\label{scurrent}
{\cal J}_{\pm}= -i G^{-1}D_{\pm}G
\eea
This current is associated with $SU(2)_R$ and there is another supercurrent
associated with $SU(2)_L$. The existence of such currents follow from 
invariance of the action under global symmetries as mentioned earlier. 
Furthermore, from the structure of the current (\ref{scurrent}), we conclude
that they also satisfy a curvaturelessness condition in the superspace:
$D_-{\cal J}_++D_-{\cal J}_+ +i[{\cal J}_+ ,{\cal J}_-]_+=0 $; 
here $[,]_+$ stands
for anticommutator of the currents. 
If we are not at the superconformal value of
the couplings i.e. ${\lambda ^2 k\over {4\pi}} \ne \pm1$, 
then the conservation law
does not have holomorphic property i.e. the conservation equations are 
$(1-\delta)D_+{\cal J}_- -(1+\delta )D_-{\cal J}_+ =0$ and $\delta =\pm 1$
is the critical coupling constant values to ensure conformal invariance.
It is obvious that at this critical value, the equation of motion involves
only one current since the other term vanishes identically, exhibiting
the holomorphicity property. There is another current which is 
anti-holomorphic.
When we choose
$\delta =1$,
 the equation of motion becomes
\be
D_-{\cal J}_+ =0
\ee
This is supersymmetric version of a holomorhicity condition. In fact the
existence of holomorphic currents in such superconformal models, defined
on group manifolds, is the raison de etr\'e for the presence of underlying
Kac-Moody algebras \cite{paolo}.
We may conclude
that ${\cal J}_+$ depends only on $\sigma ^+$ and $\theta ^+$. We may expand
it in superfields (the first component being the fermion)
\be
{\cal J}_+= \psi _+ +\theta _+J_+
\ee 
The conservation law as a matrix equation is: $\partial _-J_+=0$ 
for the bosonic current, where
$J_+= -g^{-1}\partial _+g -i \psi _+ ^2$. Recall that $g$ are $\psi _+$ are
$SU(2)$ matrices (\ref{superf}) and bilinear in fermion is to be understood as
multiplication of matrix elements.\\
We proceed to construct nonlocal currents for the case under study having 
obtained the supercurrents.  Let us briefly recapitulate the two basic
ingredients for deriving the infinite family of nonlocal currents in
 non-supersymmetric chiral models \cite{chiral} without WZ term. We can 
construct
\be
{\cal A}=g^{-1}\partial _{\alpha}g
\ee
from (6)  when last term is absent. Then, according to 
the prescription of 
\cite{chiral}, one construct a covariant derivative ${\bf D}_{\alpha}
^{ij}=\delta ^{ij} \partial _{\alpha}  +{\cal A}_{\alpha}^{ij}$. It is easy
to check that the covariant derivative satisfies curvaturelessness condition
since ${\cal A}_{\alpha}$ is a pure gauge. It follows that 
 $[\partial ^{\alpha},{\bf D}_{\alpha}]=0$. The zeroth current
is identified as ${\cal A}_{\alpha}$; and then \cite{chiral} lay down the
procedure to construct the family of currents by utilizing the two properties
mentioned above. The first nonlocal conserved charge for the $\sigma$-model
is
\bea
Q_{\sigma}=\int _{-\infty}^{+\infty} dx{\cal A}_0(x,t)\int _{-\infty} ^x
{\cal A}_0(y,t)+2\int _{-\infty} ^{\infty} {\cal A}_1(x,t)
\eea
Here ${\cal A}_0 ~{\rm and }~{\cal A}_1$ correspond to the zeroth and
first (space) component of the current. The construction of nonlocal
currents for bosonic $\sigma$-model with WZ terms were presented in ref
\cite{vega,abd}.\\
In case of superchiral model, defined on a group manifold, with the WZ
term, the procedure \cite{chau,sgroup} is a generalization of 
\cite{vega,abd}. A covariant derivative is introduced in the superspace 
\bea
\label{scov}
\nabla _{\pm}=(1\pm\delta)(D_{\pm}+i[{\cal J}_{\pm},])
\eea
according to the prescription of \cite{chau,sgroup}. 
The bracket $[{\cal J}_{\pm},]$ in the above equation is to be understood
as follows: if the supercovariant derivative acts on any object which is
in $SU(2)$ and is bosonic, we take the commutator. Whereas for 
a fermionic object we take the anticommutator. It is easy to  show, 
using the definition of $D_{\pm}$ that
the anticommutator of $\nabla _+$ and $\nabla _-$ vanish;
i.e. $ \{ \nabla _+ ,\nabla _- \}=0$,  even if we are away
from the point $\delta \ne \pm 1$. The techniques of \cite{chiral,vega,abd}
for the supersymmetric WZW model is \cite{chau,sgroup} construct the
$n^{th}$ current ${\cal J}^{(n)}$ from a scalar superfield ${\cal F}^{(n)}$
through the definition
\be
{\cal J}^{(n)}_{\pm}= \pm D_{\pm} {\cal F}^{(n)}
\ee
Thus by construction ${\cal J}^{(n)}$ is conserved: $D_-{\cal J}^{(n)}_+ -
D_+{\cal J}_-^{(n)}=0$; where one utilizes the relation:  $\{D_{\pm} ,
\nabla _{\mp} \}=0 $ and definition of ${\cal J}^{(n+1)}$ in terms of
${\cal F}_{\pm}^{(n)}$ with the action of $|nabla _{\pm}$. 
In analogy with \cite{chiral} it is convenient to define
\be
\label{zeroc}
{\cal J}^0_{\pm}=(1\pm \delta){\cal J}_{\pm}
\ee
The first nonlocal current ${\cal J}^{(1)}_+ $ is given by
\be
{\cal J}^{(1)}_+=(1\pm \delta)(D_{\pm}{\cal F}^{(0)}-
{1\over 2}[{\cal J}_{\pm}, {\cal F}^{(0)}])
\ee
Let us recall some of the subtleties in construction of the nonlocal
currents for strings in the AdS background.
 In the context of two dimensional nonlinear $\sigma$-models,
 the nonlocal currents are derived by the iteration procedure
\cite{chiral,sgroup} and one has to integrate over the spatial coordinate
$x$ which extends from $-\infty $ to $+\infty$. Furthermore, as we construct
higher and higher level of currents there are ordering in the integrals
of the spatial coordinates ($\theta$-functions appear in spatial integrations).
When we deal with string theory, the $\sigma$-variable is periodic. Thus
the ordering problem in defining the integrals is to be kept in mind as has been
emphasized by Dolan, Nappi and Witten \cite{dnw}. Let
us explicitly construct the first nonlocal current. For the case at hand,
we are interested in supersymmetric WZW model for the value of coupling
where conformal invariance is guaranteed i.e. we choose the point
$\delta =1$.   Let us define
the current $J^a= -{\rm Tr}T^aJ_+$, from the bosonic component (that includes 
fermion bilinear too) of ${\cal J}_+$. Here $T^a$ are (antihermitian) 
generators of $SU(2)$ and $\epsilon ^{abc}$ being the structure constants. 
The zeroth component of the  first nonlocal current  is
\bea
\label{firstcr}
J^{(1)a}_0=J^{(0)a}_1(\sigma)+J^{(0)a}_0(\sigma)-i\epsilon ^{abc}\psi ^b_+
\psi ^c_+ -{1\over 2}\epsilon ^{abc} \int ^{\sigma}d\sigma '
J^{(0)b}_0(\sigma)J^{(0)c}_0
(\sigma ') 
\eea
and the corresponding (first) nonlocal charge is the $\sigma$ integration of
the above expression. Thus, in the last term, one has to integrate over
the  $\sigma '$ variable. Notice that the first nonlocal charge
is integral over $\sigma  $  which is periodic. Therefore, the
last term will have integrations over $\sigma$ and $\sigma '$.  
It is obvious  that, since $\sigma$ is periodic,
strictly speaking, we do not have the notion of $\sigma ' <\sigma$ in
the integrals of the last term in defining the charge. The issues related
to the definition of charges and periodic boundary conditions have been
addressed by Dolan and Nappi \cite{dn}. 
Of course, this situation arises in all discussions concerning
the definition of nonlocal (currents) charges in the context of closed
strings in $AdS_3 \times S^3$ or $AdS_5 \otimes S^5$ backgrounds.\\
Therefore, there exist an infinite family of nonlocal conserved currents
for NS-NS string propagating on $S^3$, although there are technical issues
associated with the periodicity of $\sigma$ as remarked earlier.\\
Now let us turn our attention to the $AdS_3$ space. The group manifold is
$SL(2,R)$ and it is a noncompact group. Therefore, the coefficient of
WZ term is not quantised as mentioned before. The construction of the
action achieved in a similar manner and one has to take into account following
aspects while proceeding. (i) Since the bosonic matrix field 
${\tilde g}\in SL(2,R) $ satisfies
${\tilde g}^T\eta {\tilde g}=\eta$, the corresponding superfield, $\tilde G$ is 
required to fulfil the constraint ${\tilde G}^T\eta {\tilde G} =\bf 1$.
Then, we can obtain the corresponding supercurrents which will satisfy
conservation law. The requirement of superconformal invariance will enforce
corresponding equations of motion (current conservation) to be superholomorphic
as was the case with $SU(2)$. Subsequently, we can construct the family of
nonlocal currents for the $AdS_3$ background as we have done for the $SU(2)$
geometry.\\
The existence an infinite family of conserved currents can lead to Ward
identities. We shall outline, very briefly, the prescriptions to derive
these Ward identities. Let us consider the classical situation in the sense that
we ignore the presence of any anomalies which might vitiate the
identities in the full quantum computations. The starting point is to consider
the expectation of product of a set of vertex operators, $V(Y)$, 
where $Y(\tau , \sigma)$
are the string coordinates defining the $S^3$.
\bea
\label{wi}
<V(Y_1)V(Y_2)...V(Y_n)>
\eea
Here $<....> $ is defined with the generating functional ${\bf Z}=
\int  [{\cal D}Y]
e^{iS[Y]}$ and \\
$Y_1(\tau _1,\sigma _1), Y_2(\tau _1, \sigma _1),....$ are 
string coordinates on $S^3$ and $S[Y]$ is the corresponding string action.. 
These vertex operators satisfy the requirements of conformal
invariance. We can compute a correlation function with one of the nonlocal
currents, say $J^{(1)}_{\alpha}$
\bea
\label{corr}
<J_{\alpha}^{(1)}(Y(\tau , \sigma)V(Y_1)V(Y_2)....V(Y_n)>
\eea
We can compute the above expression which involves the current. When we
use the current conservation,
\bea
\partial _{\alpha}<J_{\alpha}^{(1)}(\tau ,\sigma)V(Y_1)V(Y_2)...V(Y_n)> =0
\eea
Note that $\partial _{\alpha}$ acts on the current which depends
on $(\sigma , \tau)$ through its dependence on the string coordinates.
This is the Ward identity. As an illustrative example, consider the WI
involving graviton vertex operators; again for the $S^3$ case focusing on the
bosonic part alone. 
The vertex operator is defined to be
\bea
\label{vo}
{V_{ij}(Y_n)}={\delta \over {\delta g_{ij}(Y_n(\tau _n,\sigma _n))}}S[Y]
\eea
If we eliminate extra coordinate $Y_4$ (see (\ref{su2})), then the 
target space metric (for$S^3$)  is given by
\bea
g_{ij} =\delta _{ij} +{ Y_i Y_j \over {(1-\sum Y_iY_i)^{{1\over 2}}}}
\eea
The WI can be obtained following the prescription of Maharana and Veneziano
\cite{jmgv}, since for the bosonic case the nonlocal current
$J_{\alpha}^{(1)}$ involves the spatial component of the zeroth current
and product of the time components of the zeroth current as can be read off from
(\ref{firstcr}) by setting the fermion fields to zero. Note that the in 
establishing quantum WI several points are to be taken into account. First of
all the nonlocal current, being defines as products of lower currents, 
have to be defined with an appropriate prescription. Second, the issue 
related to the periodicity of $\sigma$ is to be addressed. Finally, if the
current conservation law is afflicted with anomalies \cite{ml} that should be 
reflected in the WI. It is worth while to mention that the infinite
sequence of nonlocal currents are conserved on shell i.e. the conservation
laws are valid when the field equations are satisfied.\\
It
will be interesting to examine, if existence of these charges have
any consequences in understanding the $AdS_3 /{CFT_2}$ correspondence 
\cite{mald,givn,bars}.
This is a pertinent issue  since
there is interesting connection between  integrability of type IIB theory 
 on $AdS_5 \otimes S^5$ and the spin chain approach to study anomalous 
dimension of some operators in ${\cal N} =4$ SYM theories. Moreover, the
WI we propose might be useful to further explore the consequences of
$AdS_3/CFT_2 $.\\
In summary, we have considered NSR string in the 10-dimensional spacetime, 
in the background geometry, $AdS_3 \otimes S^3
\otimes R_4$. There are covariantly constant 3-form fluxes along $AdS_3$ and
$S^3$ direction and it is set to zero along the flat Euclidean space $R_4$.
It is argued that the worldsheet action for $AdS_3 \otimes S^3$ geometry along
with the fluxes can be described as super symmetric WZW model
 on the group manifolds $SL(2,R) \otimes SU(2)$. We present the construction of
an infinite sequence of nonlocal conserved currents and point out the
difficulties due to the periodicity of $\sigma$ which is inherent in the
previous constructions. We argue that these conservation laws lead to
Ward identities for the correlation functions of vertex operators. We outline
the derivation of such WI for the classical case. It is conjectured that
these WI's might be utilized as useful tools in the study of $AdS_3$/ $CFT_2$
correspondence.

We note that a family of nonlocal conserved currents are shown to exist
for a ten dimensional superstring with NS-NS fermions in the background
geometry of $AdS_3 \otimes S^3 \otimes R_4$. The field strength of
antisymmetric tensor is constant on $AdS_3$ as well as on $S^3$, whereas it
is set to zero on $R_4$. The currents, we presented here
are classical ones, modulo the problem associated with the periodicity of
$\sigma$. It is worth examining the question whether the  conservation laws are
free from anomalies in the quantum theory \cite{ml}.

It is well known that string theories possess rich symmetry structures. 
These symmetries have played central role in our understandings of
various facets of string theories. 

\noindent {\bf Acknowledgments:} It is a pleasure to thank John Schwarz for
very valuable discussions on strings in $AdS_3$ backgrounds and their
integrability properties. I have benefited from discussions
with  Rajesh Gopakumar.

\newpage




\end{document}